\documentclass[fleqn,a4paper]{vch-book}
\usepackage{amsmath}
\usepackage{amssymb}
\usepackage{makeidx}
\usepackage{latexsym}
    \makeindex
\usepackage{times}
\usepackage{tabularx}
\usepackage{booktabs}
\usepackage[dvips]{epsfig}
\DeclareGraphicsExtensions{.eps,.ps}
\chardef\bslash=`\\ % p. 424, TeXbook
\newcommand{\ntt}{\normalfont\ttfamily}
\newcommand{\bibtex}{\ifx\is@itshape\f@shape{\fontshape{scit}\selectfont
Bib}\else\textsc{Bib}\fi\kern-.1em\TeX}

\newcommand{\kcm}{(k_c / m)}
\newcommand{\equ}[1]{(\protect\ref{#1})}
\newcommand{\avk}{\left< k \right>}
\newcommand{\fluck}{\left< k^2 \right>}

\begin{document}

\author{R. Pastor-Satorras and A. Vespignani}
\title{Epidemics and immunization in scale-free networks}

\maketitle
%\tableofcontents

\chapter{Epidemics and immunization in scale-free networks \for{toc}{\newline}
  \tocauthor{Romualdo Pastor-Satorras and Alessandro Vespignani}}

\authorafterheading{Romualdo Pastor-Satorras}

\affil{Departament  de F{\'\i}sica i Enginyeria Nuclear\\
  Universitat Polit{\`e}cnica de Catalunya\\
  08034 Barcelona, Spain}
\vspace*{1.5cm}
\authorafterheading{Alessandro Vespignani}

\affil{The Abdus Salam International Centre for Theoretical Physics
  (ICTP)\\
  P.O. Box 586, 34100 Trieste, Italy}

\section{Introduction}
\label{sec:intro}

Epidemic models are heavily affected by the connectivity patterns
characterizing the population in which the infective agent
spreads~\cite{bailey,epidemics,anderson92}.  Many models map this
pattern in terms of networks, in which nodes represent individuals and
links represent the possible contacts along which the epidemic
diffuses.  In this perspective scale-free (SF) networks
\cite{barab99,mendes99,barabasi01} represents a very interesting cases
since they exhibit a power-law connectivity distribution
\begin{equation}
  P(k)\sim k^{-\gamma}
\end{equation}
for the probability $P(k)$ that a node of the network has $k$
connections to other nodes. For connectivity exponents in the range $2
< \gamma \leq 3$ this fact implies that each node (element of the
population) has a statistically significant probability of having a
very large number of connections compared to the average connectivity
$\avk$ of the network. In mathematical terms, the implicit divergence
of $\fluck$ is signalling the extreme heterogeneity of the
connectivity pattern, and it easy to foresee that this property is
going to change drastically the behavior of epidemic outbreaks in SF
networks.  The interest in the study of epidemic models in SF networks
is enhanced by the evidence that both the
Internet~\cite{falou99,calda00,alexei,yook01,goh01b} and the maps of
human sexual contacts~\cite{amaral01} are characterized by scale-free
connectivity properties. The Internet and the web of sexual contacts
are, in fact, the natural environment in which cyber viruses and
sexually transmitted diseases (STD), respectively, live and
proliferate.  The study and characterization of epidemics in
scale-free networks is therefore of potential importance for those
seeking to control and arrest human and electronic plagues. These
considerations motivate the analysis of the effect of complex network
topologies in standard epidemic models leading to several interesting
and novel results \cite{abramson01,moore00,pv01a}.

Perhaps, the most surprising result, first originated by the
inspection of the susceptible-infected-susceptible (SIS) model, is
that the spread of infections is tremendously strengthened on SF
networks~\cite{pv01a,pv01b}. Opposite to standard models, epidemic
processes in these networks do not possess any epidemic threshold
below which the infection cannot produce a major epidemic outbreak or
an endemic state. In principle, SF networks are prone to the
persistence of diseases whatever infective rate they may have. The
same peculiar absence of an epidemic threshold has been readily
confirmed in other epidemic models such as the
susceptible-infected-removed (SIR)
model~\cite{lloydsir,moreno,newman02} and appears as a general feature
of epidemic spreading in SF networks.  This feature reverberates also
in the choice of immunization strategies~\cite{psvpro,aidsbar} and
changes radically the standard epidemiological framework usually
adopted in the description and characterization of disease
propagation.

In this chapter we want to provide a review of the main results
obtained in the modeling of epidemic spreading in SF networks.  In
particular, we want to show the different epidemiological framework
originated by the lack of any epidemic threshold and how this feature
is rooted in the extreme heterogeneity of the SF networks'
connectivity pattern. As a real world example of epidemic spreading
occurring on SF connectivity patterns, we shall consider the diffusion
of computer viruses. Computer virus spreading, in fact, can be
characterized by simple population models that do not consider
properties such as gender, sex, or age, that must be included in
the modeling of STD and other kinds of epidemics. On the other hand,
computer viruses proliferate in the Internet, that is a capital
example of SF network, and it is natural to include this topology in
their modeling.  Finally, many real data are available in computers
epidemiology and we can use them to show experimentally the failure of
the standard epidemic framework and support the new picture arising
for SF networks.

We also present how the scale-free nature of the network calls for
different immunization strategies in order to eradicate infections.
Opposite to standard models, it is found that SF networks do not
acquire global immunity from major epidemic outbreaks even in the
presence of unrealistically high densities of randomly immunized
individuals.  Successful immunization strategies, therefore, can be
developed only by taking advantage of the inhomogeneous connectivity
properties of the scale-free connectivity patterns.  Finally we
consider the effect of the network finite size, referring to real
systems which are actually made up by a finite number of individuals.
The presented results provide a general view of the novel features of
epidemic models in SF networks that, besides the application to
computer viruses, prompt to the relevant implications of these studies
in human and animal epidemiology.

\section{Computers and epidemiology}
\label{sec:computer-viruses:-an}

In a classic paper \cite{bellovin}, it is described the
Domain-Name-Server (DNS) cache corruption spreading as a
\textit{natural computer virus} proliferating on the Internet.
Computers on the Internet rely upon DNS servers to translate Internet
protocol addresses into computer names and vice-versa.  On their turn,
DNS servers communicate with their DNS peers to share and update these
informations. The updating is periodic in time and in the meanwhile,
translation tables are ``cached'' and eventually transmitted to the
other DNS peers. If any portion of this cache is corrupted, the DNS
server will provide incorrect addresses not only to requesting
computers but to DNS peers as well, propagating the error. At the same
time, any DNS server can get ``cured'' by an updating with an
error-free DNS peer.  The same kind of processes can occur with
routing tables exchanged by routers. This propagation of errors
occurring on routers and servers that are physically linked is a
typical example of epidemic process, in which the corruption (virus)
is transmitted from infected to healthy individuals.

From a more familiar point of view, however, computer viruses are
usually referred to as little programs that can reproduce themselves
by infecting other programs \cite{virusgeneral,scientific97}. The
basic mechanism of infection is as follows: When the virus is active
inside the computer, it is able to copy itself, by different ways,
into the code of other, clean, programs. When the newly infected
program is run into another computer, the code of the virus is
executed first, becoming active and being able to infect other
programs.  Apart from reproducing themselves, computer viruses perform
threatening tasks that range from flashing innocuous messages on the
screen to seriously corrupt data stored in the computer.
These deleterious effects render most computer viruses as dangerous as
their biological homonyms, and explain the interest, both commercial
and scientific, arisen around their study.

Computer viruses have evolved in time (driven of course by their
programmers' skills), adopting different strategies that take
advantage of the different weak points of computers and software.
Computer viruses can be classified into three main classes, or
\textit{strains}~\cite{scientific97}. The first strain includes
\textit{file viruses} that infect application programs.  A second and
more harming family contains the \textit{boot-sector viruses} that
infect the boot sector of floppy disks and hard drives, a portion of
the disk containing a small program in charge of loading the operating
system of the computer.  A third and nowadays prevailing strain is
formed by the \textit{macro viruses}. These viruses are independent of
the platform's hardware and infect data files, such as documents
produced with spreadsheets or word processors. They are coded using
the \textit{macro} instructions that are appended in the document,
instructions used to perform a set of automatic actions, such as
formatting the documents or typing long sequences of characters.  In
addition, with the ever more efficient deployment of antivirus
software, more harmful viruses combining together the properties of
the main strains have been developed.

Noticeably, however, the nowadays dominant and most aggressive type of
cyber organisms is represented by the \textit{worms} family.  Worms
are actually viruses infecting the computer with mechanisms similar to
usual viruses and making a particularly effective use of the e-mail
for infecting new computers. In fact, by using the instructions of
some commercial mail software applications, worms are capable of sending
themselves to all the e-addresses found in the address-book of the
person receiving the infected mail. This possibility renders worms the
most effective viruses, especially in terms of the velocity at which
they can propagate starting from a single infection.

The spreading of computer viruses has been studied for long years, in
close analogy with the models developed for the study of the
transmission of biological diseases (for a review see
Refs.~\cite{kep91,ieee93}). In this biological framework, the key
point is the description of the epidemic process in terms of
\textit{individuals} and their \textit{interactions}. In this
simplified formalism, individuals can only exist in a discrete set of
states, such as susceptible (or healthy), infected (and ready to
spread the disease), immune, dead (or removed), etc. On the other
hand, the interactions among individuals are schematized in the
structure of the contacts along which the epidemics can propagate.
Within this formalism, the system can be described as a
\textit{network} or graph \cite{chartrand}, in which the nodes
represent the individuals and the links are the connections along
which the epidemics propagates.

Standard epidemiological models usually consider \textit{homogeneous}
networks, which are those that have a connectivity distribution peaked
at an average connectivity $\avk$, and decaying exponentially fast for
$k\ll\avk$ and $k\gg\avk$.  A typical example of deterministic
homogeneous network is the standard hypercubic lattice, while among
the random homogeneous network we can count the Erd\"{o}s-R{\'e}nyi model
\cite{erdos60} and the Watts-Strogatz model \cite{watts98}.  On the
other hand, as we shall see in the following, computer viruses and
worms spread in environments characterized by scale-free
connectivities. This will lead to the failure of the standard epidemic
picture and will naturally introduce the scale-free connectivity as an
essential ingredient for the understanding of computer viruses.

\section{Epidemic spreading in homogeneous networks}
\label{sec:sis-model-homogenous}

The simplest epidemiological model one can consider is the
susceptible-infected-susceptible (SIS) model
\cite{epidemics,anderson92}.  In the SIS model, individuals can only
exist in two discrete states, namely, susceptible and infected.  These
states completely neglect the details of the infection mechanism
within each individual.  The disease transmission is also described in
an effective way.  At each time step, each susceptible node is
infected with probability $\nu$ if it is connected to one or more
infected nodes. At the same time, infected nodes are cured and become
again susceptible with probability $\delta$, defining an effective
\textit{spreading rate}
\begin{equation}
  \label{eq:deflambda}
  \lambda=\frac{\nu}{\delta}.
\end{equation}
Without lack of generality, we set can set $\delta=1$, since it only
affects at the definition of the time scale of the disease
propagation.  Individuals thus run stochastically through the cycle
\begin{center}
  susceptible $\to$ infected $\to$ susceptible,
\end{center}
and hence the name of the model.  The SIS model does not take into
account the possibility of individuals removal due to death or
acquired immunization which would lead to the so-called
susceptible-infected-removed (SIR) model \cite{epidemics,anderson92}.
It is mainly used as a paradigmatic model for the study of infectious
disease leading to an endemic state with a stationary and constant
value for the density of infected individuals, i.e. the degree to
which the infection is widespread in the population. The SIS has been
adopted in the modeling of computer viruses and worms since, also in
the presence of antiviruses, computer immunization 
statistically depends upon the user concerns in not skipping the antivirus
control when opening e-mail attachments or new files.
 
The analytical study of the SIS model can be undertaken in terms of a
dynamical mean-field (MF) theory.  For homogeneous networks, in which
the connectivity fluctuations are very small, we can approach the MF
theory by means of a reaction equation for the total prevalence
$\rho(t)$, defined as the density of infected nodes present at time $t$.
That is, we can consider all the nodes as equivalent, irrespective of
their corresponding connectivity. The reaction equation for $\rho(t)$
can be written as
\begin{equation}
  \partial_t \rho(t) = -\rho(t) +\lambda \avk \rho(t) \left[ 1-\rho(t) \right].
\label{eq:ws}
\end{equation}
The MF character of this equation stems from the fact that we have
neglected the density correlations among the different nodes.  In
Eq.~\equ{eq:ws} we have also ignored all higher order corrections in
$\rho(t)$, since we are interested in the onset of the infection close
to the point $\rho(t) \ll 1$.  The first term on the right-hand-side in
Eq.~\equ{eq:ws} considers infected nodes becoming healthy with unit
rate. The second term represents the average density of newly infected
nodes generated by each active node. This is proportional to the
infection spreading rate, $\lambda$, to the number of links emanating from
each node, and to the probability that a given link points to a
healthy node, $\left[ 1-\rho(t) \right]$.  In the homogeneous networks
we are considering here, connectivity has only very small fluctuations
($\fluck \sim \avk$) and as a first approximation we have considered
that each node has the same number of links, $k\simeq \avk$.  This is
equivalent to an homogeneity assumption for the system's connectivity.
In writing this last term of the equation we are also assuming the
\textit{homogeneous mixing hypothesis} \cite{anderson92}, which
asserts that the force of the infection (the per capita rate of
acquisition of the disease for the susceptible individuals) is
proportional to the density of infected individuals $\rho(t)$. The
homogeneous mixing hypothesis is indeed equivalent to a mean-field
treatment of the model, in which one assumes that the rate of contacts
between infectious and susceptibles is constant, and independent of
any possible source of heterogeneity present in the system.  Another
implicit assumption of this model is that the time scale of the
disease is much smaller than the lifespan of individuals; therefore we
do not include in the equations terms accounting for the birth or
natural death of individuals.

\begin{vchfigure}[t]
    \epsfig{file=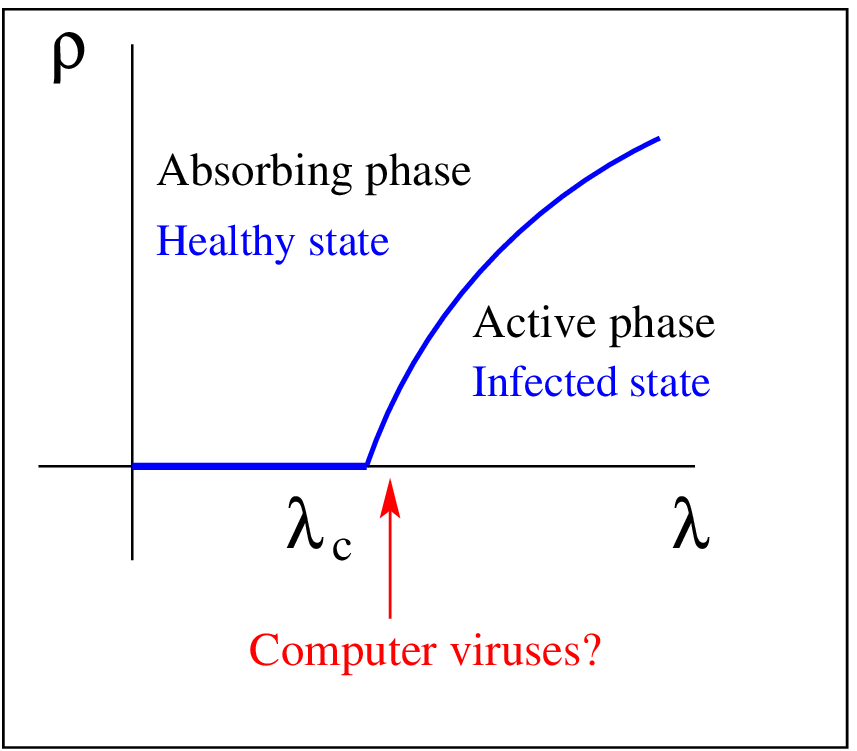, width=6cm}
    
    \vchcaption{Schematic phase diagram for the SIS model in
      homogeneous networks. The epidemic threshold $\lambda_c$ separates an
      active or infected phase, with finite prevalence, from an
      absorbing or healthy phase, with null prevalence. The very small
      prevalence and long lifetimes observed in computer virus data
      are only compatible with a value of $\lambda$ infinitesimally close
      to the epidemic threshold.}
  \label{fig:diagram}
\end{vchfigure}

After imposing the stationarity condition $\partial_t \rho(t) =0$, we obtain
the equation, valid for the behavior of the system at large times,
\begin{equation}
  \rho \left[ -1 + \lambda \avk (1-\rho) \right]=0
\end{equation}
for the steady state density $\rho$ of infected nodes.  This equation
defines an epidemic threshold $\lambda_c=\avk^{-1}$, and yields:
\begin{eqnarray}
  \rho &=& 0 \,\qquad\qquad \qquad\mbox{\rm if $\lambda< \lambda_c$}, \\
  \rho &=& (\lambda-\lambda_c)/ \lambda  \qquad \mbox{\rm if $\lambda\geq \lambda_c$} \label{eq:meanfield}.
\end{eqnarray}

The most significant prediction of this model is the presence of a
nonzero \textit{epidemic threshold} $\lambda_c$ \cite{marro99,epidemics}.
If the value of $\lambda$ is above the threshold, $\lambda\geq \lambda_c$, the
infection spreads and becomes persistent.  Below the threshold, $\lambda<
\lambda_c$, the infection dies exponentially fast. From the point of view
of nonequilibrium phase transitions, the SIS model exhibits an
\textit{absorbing-state phase transition} \cite{marro99} at the
threshold $\lambda_c$, separating an active or infected phase, with finite
prevalence, from an absorbing or healthy phase, with null prevalence.
A qualitative picture of the phase diagram of this transition is
depicted in Figure~\ref{fig:diagram}.  It is easy to recognize that
the SIS model is a generalization of the contact process model, widely
studied in this context as the paradigmatic example of an
absorbing-state phase transitions to a unique absorbing state
\cite{marro99}.

To summarize, the main prediction of the SIS model in homogeneous
networks is the presence of a \textit{positive} epidemic threshold,
proportional to the inverse of the average number of neighbors of
every node, $\avk$, below which the epidemics always dies, and endemic
states are impossible.

\section{Real data analysis} 
\label{sec:stat-prop-comp}

The statistical properties of computer virus data have been analyzed
by several authors, in close analogy with the classical epidemiology
of biological diseases \cite{bailey,epidemics,anderson92}.  Within
this framework, studies have focused specially in the measurement of
the virus \textit{prevalence}, defined as the average fraction of
computers infected with respect to the total number of computers
present. From these studies \cite{scientific97,SSP93*2,ieee93} two
main conclusions have been drawn. First, viruses which are able to
survive in order to produce a significant outbreak usually reach an
\textit{endemic} or metastable steady state, with a stationary
prevalence.  The second empirical observation is that these endemic
viruses do attain in general a very small average prevalence, that can
be of the order of one out of $1 000$ computers or less.

\begin{vchfigure}[t]
    \epsfig{file=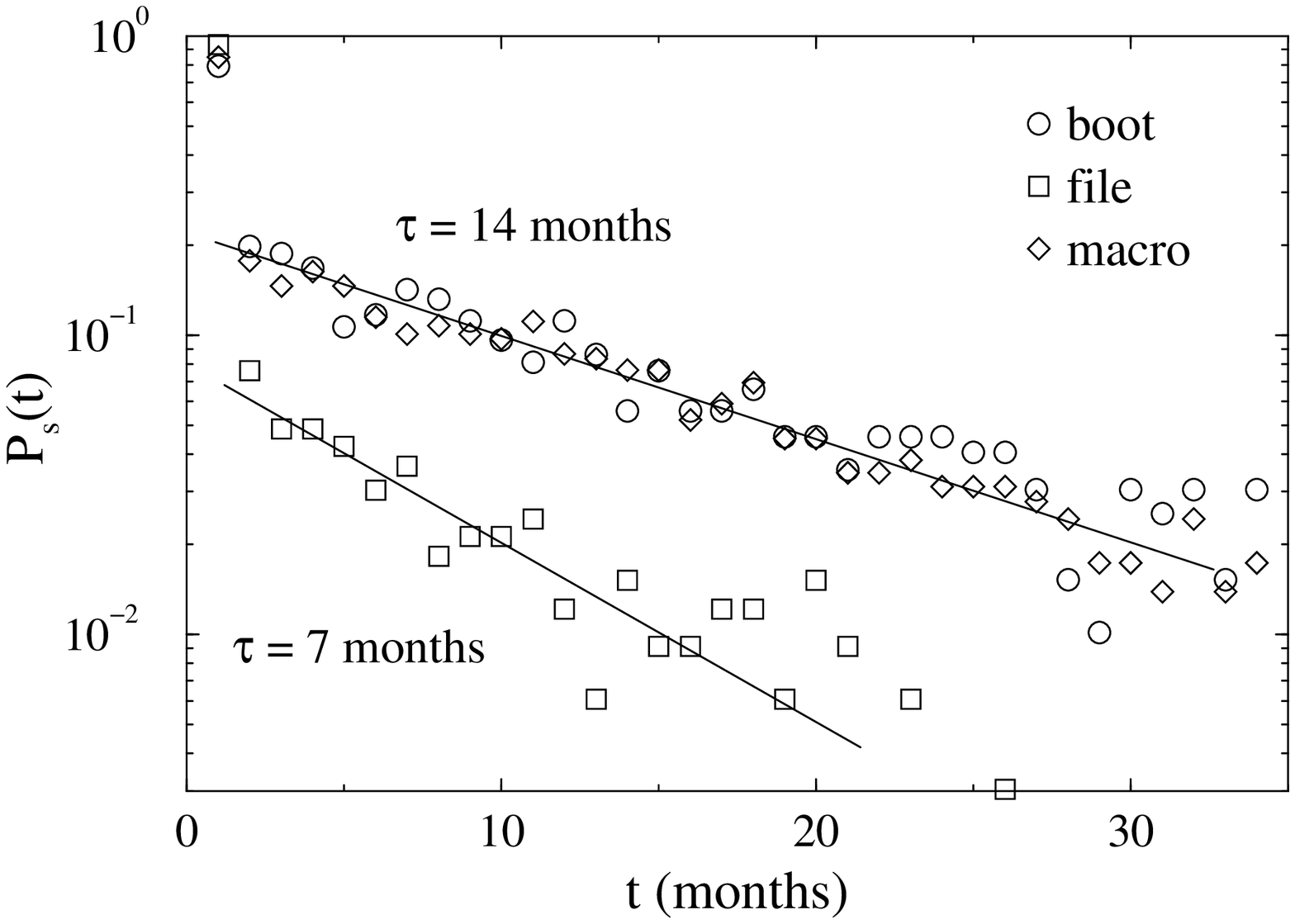, width=8cm}
    
    \vchcaption{Surviving probability for the three main strands of
      computer viruses. After a sharp initial drop, it is clear the
      presence of an exponential decay, with an associated
      characteristic time $\tau$ that depends on the given strand.}
  \label{fig:surviving}
\end{vchfigure}

More recently, other studies \cite{pv01a} have focused in the dynamics
of the spreading process, measuring the {\em surviving probability} of
homogeneous groups of viruses, classified according to their infection
mechanism (strains). In these studies one considers the total number
of viruses of a given strain that are born and die within a given
observation window.  The surviving probability $P_s(t)$ of the strain
is defined as the fraction of viruses still alive at time $t$ after
their birth. Figure~\ref{fig:surviving} reproduces the results
reported in Ref.~\cite{pv01a}, obtained from prevalence data from the
\textit{Virus Bulletin}\footnote{Virus prevalence data publicly
  available at the web site {\ntt http://www.virusbtn.com}.}  in the
period February 1996 to March 2000, covering a time interval of 50
months.

Figure~\ref{fig:surviving} shows that the surviving probability
suffers a sharp drop in the first two months of a virus' life.
On the other hand, Figure~\ref{fig:surviving}
also shows for larger times a clean exponential tail, 
\begin{equation}
  P_s(t)\sim\exp(-t/\tau),
\end{equation}
where $\tau$ represents the characteristic life-time of the
virus strain.  The numerical fit of the data~\cite{pv01a} yields $\tau\simeq
14$~months for boot and macro viruses and $\tau\simeq 6-9$ months for file
viruses.

When comparing the theoretical picture delivered by the SIS model on
homogeneous networks with the behavior observed in real computer
viruses, one is faced with an unexpected and paradoxical conclusion.
First of all, the extremely low prevalence shown by endemic viruses is
only compatible with the phase diagram sketched in
Figure~\ref{fig:diagram} in the very \textit{unlikely} chance that all
surviving viruses are constructed such that their respective spreading
rate $\lambda$ is tuned infinitesimally close to $\lambda_c$, above the epidemic
threshold.  On the other hand, the characteristic life times observed
in the analysis of the surviving probability of the different virus
strains are impressively large if compared with the interval in which
anti-virus software is available on the market (usually within days or
weeks after the first incident report) and corresponds to the
occurrence of metastable endemic states.  Such a long lifetime on the
scale of the typical spread/recovery rates would suggest an effective
spreading rate larger than the epidemic threshold, which is in
contradiction with the always low prevalence levels of computer
viruses but in the case of an unrealistic tuning of all viruses to the
system epidemic threshold.  In summary, the comparison with the known
experimental data points out that the view obtained so far with the
modeling of computer viruses is very instructive, but fails to
represent, even at a qualitative level, the nature of the real
phenomenon. The explanation of this discrepancy has been claimed to be
one of the most important open problems in computer virus epidemiology
\cite{white98}.

The key point to elucidating the riddle posed by computer viruses
resides in the capacity of many of them to propagate via data exchange
with communication protocols (FTP, e-mails, etc.) \cite{scientific97}.
Viruses will spread preferentially to computers which are highly
connected to the outer world and are thus proportionally exchanging
more data and information. It is thus rather intuitive to consider the
scale-free Internet connectivity as the effective one on which the
spreading occurs. For instance, this is the case of \textit{natural
  computer viruses} which spread on the topology identified by routers
and servers~\cite{falou99,calda00,alexei}.  Apart from the Internet,
scale-free properties emerge also in the
world-wide-web~\cite{barab99,www99} and in social
networks~\cite{wass94}.  The fact that all virus strains show the same
statistical features indicates that very likely all of them spread on
scale-free connectivity patterns.  Further and strong support for this
conclusion comes from the recent study of a social network of e-mail
exchange within a community of users \cite{ebel02}, which was proven
to have a scale-free connectivity, with an exponent close to $2$. This
finding has an immediate repercussion on the modeling of worms, whose
spreading environment is in fact given by this kind of network.

The conclusion from the above arguments is that computer viruses
spread in a scale-free network, in which, even though the average
connectivity is well defined, the connectivity fluctuations are
unbounded; i.e. there is always a finite probability that a node has a
number of neighbors much larger than the average value. These
fluctuations in the connectivity are the key difference with respect
to the epidemic models discussed in homogeneous graphs, and they must
be included in a correct characterization of the system.

\section{Epidemic spreading in scale-free networks}
\label{sec:epid-spre-scale}

In order to fully take into account connectivity fluctuations in a
analytical description of the SIS model, we have to relax the
homogeneity assumption used for homogeneous networks, and work instead
with the relative density $\rho_k(t)$ of infected nodes with given
connectivity $k$; i.e. the probability that a node with $k$ links is
infected. The dynamical mean-field equations can thus be written as
\cite{pv01a,pv01b}
\begin{equation}
  \frac{ d \rho_k(t)}{d t} = 
  -\rho_k(t) +\lambda k \left[1-\rho_k(t) \right] \Theta[\{\rho_k(t)\}], 
\label{mfk}
\end{equation}
where also in this case we have considered a unitary recovery rate and
neglected higher order terms ($\rho_k(t)\ll 1$).  The creation term
considers the probability that a node with $k$ links is healthy
$[1-\rho_k(t)]$ and gets the infection via a connected node.  The
probability of this last event is proportional to the infection rate
$\lambda$, the real number of connections $k$, and the probability
$\Theta[\{\rho_k(t)\}]$ that any given link points to an infected node. We
make the assumption that $\Theta$ is a function of the partial densities
of infected nodes $\{\rho_k(t)\}$.  In the steady (endemic) state, the
$\rho_k$ are functions of $\lambda$. Thus, the probability $\Theta$ becomes also
an implicit function of the spreading rate, and by imposing the
stationarity condition $\partial_t \rho_k(t) =0$, we obtain
\begin{equation}
  \rho_k=\frac{k \lambda\Theta(\lambda)}{1+k \lambda\Theta(\lambda)}.
  \label{nhom}
\end{equation}
This set of equations show that the higher the node connectivity, the
higher the probability to be in an infected state. This inhomogeneity
must be taken into account in the computation of $\Theta(\lambda)$.  The exact
calculation of $\Theta$ for a general network is a very difficult task.
However, we can exactly compute its value for the case of a
\textit{random} SF network, in which there are no correlations among
the connectivities of the different nodes \cite{pv01a,pv01b}.  Indeed,
the probability that a link points to a node with $s$ connections is
equal to $sP(s)/ \avk$, which yields an average probability of a link
pointing to an infected node
\begin{equation}
  \Theta(\lambda)=\frac{1}{\avk}\sum_k kP(k)\rho_k.
  \label{first}
\end{equation}
Since $\rho_k$ is on its turn a function of $\Theta(\lambda)$, we obtain a
self-consistency equation that allows to find $\Theta(\lambda)$ and an explicit
form for Eq.~(\ref{nhom}).  Finally, we can evaluate the order
parameter (persistence) $\rho$ using the relation
\begin{equation}
  \rho=\sum_kP(k)\rho_k.
  \label{second}
\end{equation}

The self-consistent Eqs.~(\ref{nhom}) and~(\ref{first}) can be
approximately solved, in the limit of small $\Theta$, for any scale-free
connectivity distribution \cite{pv01b}. However, we can very easily
calculate the epidemic threshold by just noticing that $\lambda_c$ is the
value of $\lambda$ above which it is possible to obtain a nonzero solution
for $\Theta$. In fact, from Eqs.~\equ{nhom} and \equ{first}, we obtain the
self-consistent relation
\begin{equation}
  \Theta = \frac{1}{\avk}  \sum_k k P(k)  \frac{\lambda k 
\Theta}{1 + \lambda k \Theta},
\label{cons}
\end{equation}
where $\Theta$ is now a function of $\lambda$ alone~\cite{pv01a,pv01b}.  The
solution $\Theta=0$ is always satisfying the consistency equation. A
non-zero stationary prevalence ($\rho_k\neq 0$) is obtained when the
right-hand-side and the left-hand-side of Eq.~(\ref{cons}), expressed
as function of $\Theta$, cross in the interval $0<\Theta\leq 1$, allowing a
nontrivial solution.  It is easy to realize that this corresponds to
the inequality
\begin{equation}
  \frac{d}{d \Theta } \left. \left( \frac{1}{\avk}  
      \sum_k k P(k)  \frac{\lambda k
        \Theta}{1 + \lambda k \Theta} 
    \right) \right|_{\Theta =0} \geq 1
  \label{eq:critpunt}
\end{equation}
being satisfied. The value of $\lambda$ yielding the equality in
Eq.~\equ{eq:critpunt} defines the critical epidemic threshold $\lambda_c$,
that is given by
\begin{equation}
  \frac{\sum_k k P(k)  \lambda_c k}{\avk}   =  
\frac{\fluck}{\avk}  \lambda_c= 1
  \quad \Rightarrow \quad  \lambda_c = \frac{\avk}{\fluck}.
\label{thr}
\end{equation}
This results implies that in SF networks with connectivity exponent
$2<\gamma\leq 3$, for which $\fluck\to\infty$ in the limit of a network of
infinite size, we have $\lambda_c=0$.

\subsection{Analytic solution for the Barab\'{a}si-Albert network}

In order to discuss in detail a specific example, it is simpler to
consider a toy model of SF network, which is easy to generate for
simulation purposes and shows the correct connectivity properties.
The paradigmatic example of SF network is the Barab\'{a}si and Albert
(BA) model \cite{barab99,barab992,mendes99}.  The construction of the
BA graph starts from a small number $m_0$ of disconnected nodes; every
time step a new vertex is added, with $m$ links that are connected to
an old node $i$ with probability
\begin{equation}
  \Pi(k_i) = \frac{k_i}{\sum_j k_j},
\end{equation}
where $k_i$ is the connectivity of the $i$-th node.  This algorithm
implements the so-called ``rich-get-richer'' paradigm \cite{barab99},
that implies that highly connected nodes have always larger chances to
become even more connected. The networks generated this way have a
connectivity distribution $P(k) \sim k^{-3}$.

In the explicit calculations for the BA model, we use a continuous $k$
approximation that allows the practical substitution of series with
integrals \cite{barab99}.  The full connectivity distribution is thus
given by $P(k)= 2m^2k^{-3}$.  By noticing that the average
connectivity is $\avk = \int_m^\infty k P(k) d k =2m$, Eq.~(\ref{first})
gives
\begin{equation}
  \Theta(\lambda)=m\lambda\Theta(\lambda)\int_m^\infty \frac{1}{k}\frac{d k}{1+k \lambda\Theta(\lambda)} =
  m\lambda\Theta(\lambda) \log \left( 1+\frac{1}{m\lambda\Theta(\lambda)} \right),
\end{equation}
which yields the solution
\begin{equation}
  \Theta(\lambda)=\frac{e^{-1/m\lambda}}{\lambda
  m}(1-e^{-1/m\lambda})^{-1}. 
\end{equation}
In order to find the behavior of the density of infected nodes we have
to solve Eq.~(\ref{second}), that reads as 
\begin{equation}
  \rho=2m^2\lambda\Theta(\lambda)\int_m^\infty  \frac{1}{k^2}\frac{d k}{1+k \lambda\Theta(\lambda)} =
  2m^2\lambda\Theta(\lambda) \left[ \frac{1}{m} + \lambda\Theta(\lambda) \log \left(
      1+\frac{1}{m\lambda\Theta(\lambda)} \right) \right].
\end{equation}
By substituting the obtained expression for $\Theta(\lambda)$ we find at lowest
order in $\lambda$
\begin{equation}
  \rho \sim2  e^{-1/m\lambda}
\label{op}
\end{equation}

\begin{vchfigure}[t]
    \epsfig{file=sisba.eps, width=8cm}
    
    \vchcaption{Total prevalence $\rho$ for the SIS model in a BA
      network (full line) as a function of the spreading rate $\lambda$,
      compared with the theoretical prediction for a homogeneous
      network (dashed line).}
  \label{fig:sisba}
\end{vchfigure}

This result shows the absence of any epidemic threshold or critical
point in the model; i.e., $\lambda_c=0$, in agreement with the result from
Eq.~(\ref{thr}) for a scale-free network with $\fluck=\infty$. Numerical
simulations of the SIS model performed on a BA network confirm the
analytical picture extracted from the mean-field analysis.
Figure~\ref{fig:sisba} shows the total prevalence $\rho$ in the steady
state as a function of the spreading rate $\lambda$ \cite{pv01b}. As we can
observe, it approaches zero in a continuous and smooth way, compatible
with the presence of a vanishing epidemic threshold (see for
comparison the behavior expected for a homogeneous network, also drawn
in Figure~\ref{fig:sisba}). On the other hand,
Figure~\ref{fig:sisbascale} represents $\rho$ in a semilogarithmic plot
as a function of $1/ \lambda$, which shows that $\rho \sim \exp(-C /\lambda)$, where
$C$ is a constant independent of the size $N$ of the network.

\begin{vchfigure}[t]
    \epsfig{file=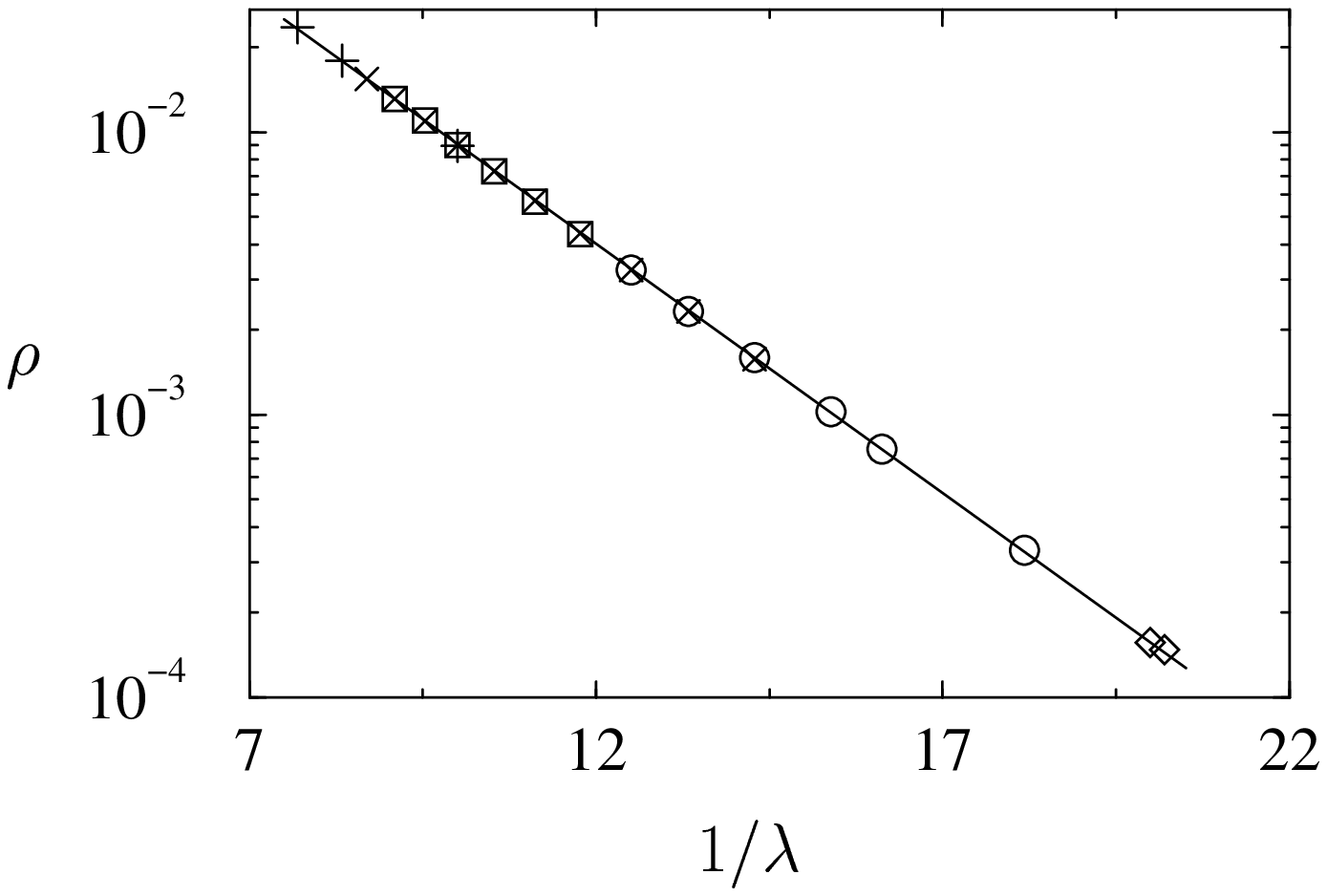, width=8cm}
    
    \vchcaption{Persistence $\rho$ as a function of $1/\lambda$ for BA
      networks of different sizes: $N=10^5$ ($+$), $N=5 \times 10^5$
      ($\Box$), $N=10^6$ ($\times$), $N=5 \times 10^6$ ($\circ$), and $N=8.5 \times 10^6$
      ($\Diamond$). The linear behavior on the semi-logarithmic scale proves
      the stretched exponential behavior predicted for the
      persistence. The full line is a fit to the form $\rho
      \sim\exp(-C/\lambda)$.}
  \label{fig:sisbascale}
\end{vchfigure}

The spreading dynamical properties of the model can also be studied by
means of numerical simulations \cite{pv01b}. For example, the
surviving probability $P_s(t)$ for a fixed value of $\lambda$ and different
network sizes $N$ is represented in Figure~\ref{fig:sisbasurv}. In
this case, we recover an exponential behavior in time, that has its
origin in the finite size of the network.  In fact, for any finite
system, the epidemic will eventually die out because there is a finite
probability that all individuals cure the infection at the same time.
This probability is decreasing with the system size and the lifetime
is infinite only in the thermodynamic limit $N\to\infty$.  However, the
lifetime becomes virtually infinite (the metastable state has a
lifetime too long for our observation window) for large enough sizes
that depend upon the spreading rate $\lambda$.  This is a well-known
feature of the survival probability in finite size absorbing-state
systems poised above the critical point \cite{marro99}.  In our case,
this picture is confirmed by numerical simulations that show that the
average lifetime of the survival probability is increasing with the
network size for all the values of $\lambda$.

\begin{vchfigure}[t]
    \epsfig{file=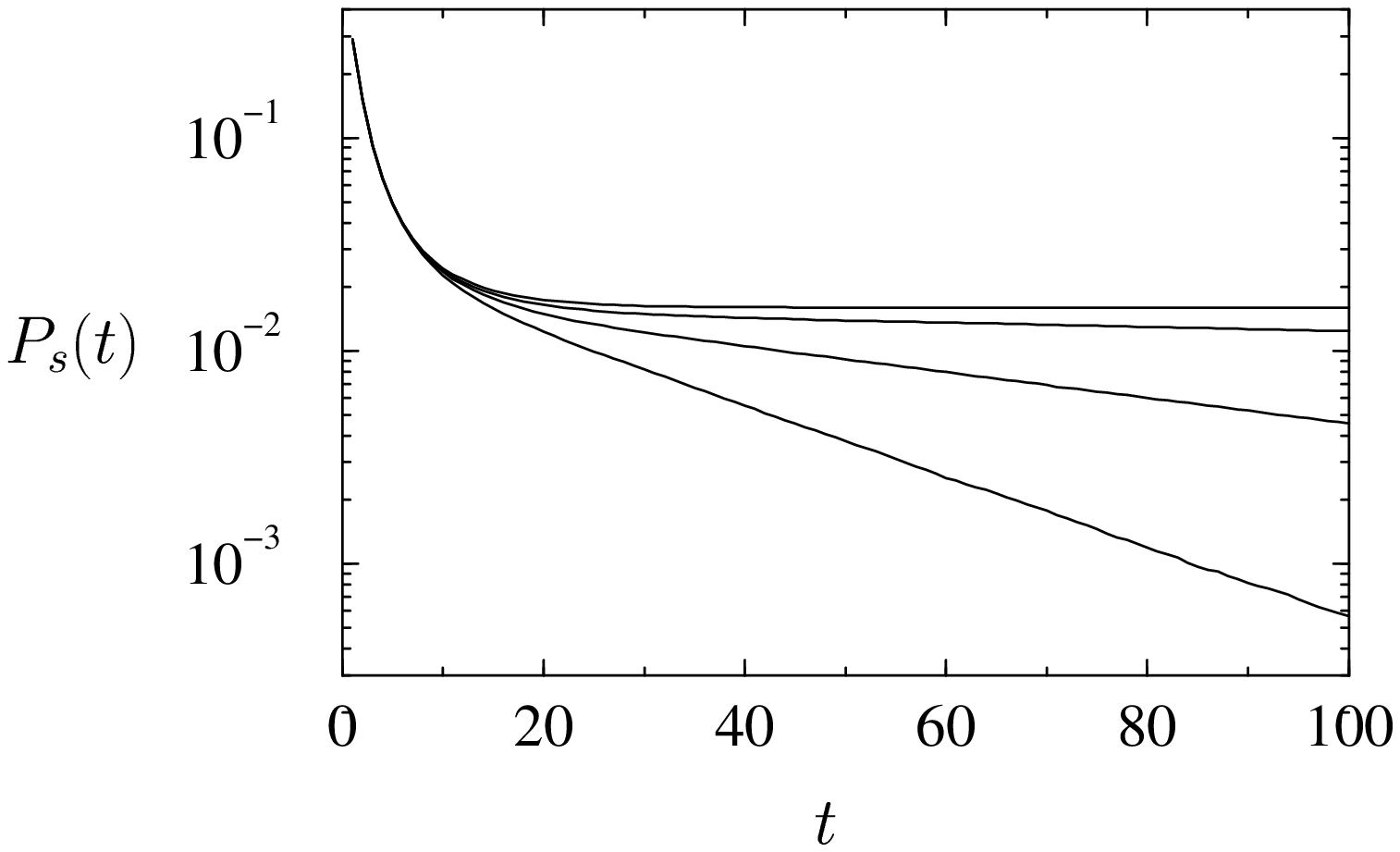, width=8cm}
    
    \vchcaption{Surviving probability $P_s(t)$ as a function of time
      in supercritical spreading experiments in the BA network.
      Spreading rate $\lambda=0.065$. Network sizes ranging from $N=6.25 \times
      10^3$ to $N=5 \times 10^5$ (bottom to top).}
  \label{fig:sisbasurv}
\end{vchfigure}

The outcome of the analysis presented in this section is that the SIS
model in a BA scale-free network, with connectivity distribution $P(k)
\sim k^{-\gamma}$ and connectivity exponent $\gamma=3$, yields the absence of
any epidemic threshold or critical point, $\lambda_c=0$. It is worth
remarking that the present framework can be generalized to networks
with $2<\gamma\leq3$, recovering qualitatively the same
results~\cite{pv01b}. Only for $\gamma>4$, 
epidemics on SF networks have the same properties as on homogeneous
networks.  The emerging picture for epidemic spreading in scale-free
networks emphasizes the role of topology in epidemic modeling. In
particular, the absence of epidemic threshold and the associated
critical behavior in a wide range of scale-free networks provide an
unexpected result that radically changes many standard conclusions on
epidemic spreading.  This indicates that infections can proliferate on
these scale-free networks whatever spreading rates they may have.
These very bad news are, however, balanced by the exponentially small
prevalence for a wide range of spreading rates ($\lambda\ll1$). This picture
fits perfectly with the observations from real data, and solve the
long-standing mystery of the generalized low prevalence of computer
viruses without assuming any global tuning of the spreading rates. In
addition, the model explains successfully the exponential time decay
of the virus surviving probability, with an average lifetime of viral
strains that appears to be related to an effective spreading rate and
the network size.

\subsection{Finite size scale-free networks}

Real systems are actually made up by a finite number of individuals
which is far from the thermodynamic limit.  This finite population
introduces a maximum connectivity $k_c$, depending on $N$, which has
the effect of restoring a bound in the connectivity fluctuations,
inducing in this way an effective nonzero threshold. More generally,
we can consider bounded scale-free networks in which the connectivity
distribution has the form $P(k) \sim k^{-\gamma} f(k/k_c)$, where the
function $f(x)$ decreases very rapidly for $x>1$
\cite{amaral,dorogorev}.  The cut-off $k_c$ can be due to the finite
size of the network or to the presence of constraints limiting the
addition of new links in an otherwise infinite networks.  In both
cases, $\fluck$ assumes a finite value in bounded SF networks,
defining from Eq.~(\ref{thr}) an effective nonzero threshold due to
finite size effects as usually encountered in nonequilibrium phase
transitions~\cite{marro99}.
\begin{vchfigure}[t]
    \epsfig{file=ratio.eps, width=8cm}
    
    \vchcaption{Ratio between the effective epidemic threshold 
    in bounded SF networks with exponential cut-off $k_c$  and the
    intrinsic epidemic threshold of homogeneous networks with the same
    average connectivity, for different values of $\gamma$.}
  \label{figkc}
\end{vchfigure}
This epidemic threshold, however, is not an {\em intrinsic} quantity
as in homogeneous systems and it vanishes for a increasing network
size or connectivity cut-off. Explicit calculations can be performed
for the SIS model \cite{pvbrief} in SF networks with exponentially
bounded connectivity, $P(k) \sim k^{-\gamma} \exp(-k/k_c)$, obtaining that
the effective nonzero epidemic threshold $\lambda_c(k_c)$ induced by the
cut-off $k_c$ behaves as
\begin{equation}
\lambda_c(k_c) \simeq \kcm^{\gamma-3},
\end{equation}
where $m$ is the smallest connectivity in the graph. The limit $\gamma \to
3$, on the other hand, corresponds to a logarithmic divergence,
yielding at leading order $\lambda_c(k_c)\simeq(m\ln(k_c/m))^{-1}$.  In all
cases we have that the epidemic threshold vanishes when increasing the
characteristic cut-off. It is thus interesting to compare the
intrinsic epidemic threshold obtained in homogeneous networks with
negligible fluctuations and the nonzero effective threshold of bounded
SF networks.  The intrinsic epidemic threshold of homogeneous networks
with constant node connectivity $\avk$ is given by $\lambda_c^{\rm
  H}=\avk^{-1}$~\cite{epidemics}.  In Fig.~\ref{figkc} we report the
ratio obtained by using the full expression for $\lambda_c(k_c)$. It is
striking to observe that, even with relatively small cut-offs ($k_c\sim
10^2-10^3$), for $\gamma\approx 2.5$ the effective epidemic threshold of finite
size SF networks is smaller by a factor close to $1/10$ than the
intrinsic threshold obtained on homogeneous networks.  This implies
that the SF networks weakness to epidemic agents is also present in
finite-size and connectivity-bounded networks.  Using the homogeneity
assumption in the case of SF networks will lead to a serious
over-estimate of the epidemic threshold even for relatively small
networks.

\section{Immunization of scale-free networks}
\label{sec:immun-compl-netw}

As we have seen in Section~\ref{sec:epid-spre-scale}, epidemic
processes in SF networks do not possess, in the limit of an infinitely
large network, an epidemic threshold below which diseases cannot set
into an endemic state. SF networks are in this sense very prone to the
spreading and persistence of infections, whatever virulence
(parametrized by the spreading rate $\lambda$) the infective agent might
possess. In view of this weakness, it becomes a major task to find
optimal immunization strategies oriented to minimize the risk of
epidemic outbreaks in SF networks.

\subsection{Uniform immunization}
\label{sec:uniform-immunization}

The simplest immunization procedure one can consider consists in the
random introduction of immune individuals in the population
\cite{anderson92}, in order to get a uniform immunization density. In
this case, for a fixed spreading rate $\lambda$, the relevant control
parameter in the density of immune nodes present in the network, the
immunity $g$. At the mean-field level, the presence of a uniform
immunity will have the effect of reducing the spreading rate $\lambda$ by a
factor $1-g$; i.e. the probability of finding and infecting a
susceptible and nonimmune node will be $\lambda(1-g)$. For homogeneous
networks we can easily see that, for a constant $\lambda$, the stationary
prevalence is given in this case by
\begin{eqnarray}
  \rho_g &=& 0 \;\,\qquad\qquad \qquad\qquad\mbox{\rm if $g>g_c$}~, \\
  \rho_g &=& (g_c-g)/(1-g)  \qquad \mbox{\rm if  $g\leq g_c$}~,
\label{eq:imm_ws1}
\end{eqnarray}
where $g_c$ is the critical immunization value above which the density
of infected individuals in the stationary state is null and depends on
$\lambda$ as
\begin{equation}
  g_c=1-\frac{\lambda_c}{\lambda}.
  \label{eq:gcWS}
\end{equation}
Thus, for a uniform immunization level larger than $g_c$, the network
is completely protected and no large epidemic outbreaks are possible.
On the contrary, uniform immunization strategies on SF networks are
totally ineffective.  The presence of uniform immunization is able to
locally depress the infection's prevalence for any value of $\lambda$, but
it does so too slowly, and it is impossible to find any critical
fraction of immunized individuals that ensures the infection
eradication. After a moment's reflection, one can convince oneself of
the reason of this failure: With the uniform immunization strategy we
are giving the same weight to very connected nodes (with the largest
infection potential) and to nodes with a very small connectivity
(which are relatively safe). Due to the large fluctuations in the
connectivity, heavily connected nodes, which are statistically very
significant, can overcome the effect of the immunization and maintain
the endemic state.  On the other hand, the absence of an epidemic
threshold ($\lambda_c=0$) in the thermodynamic limit implies that whatever
rescaling $\lambda\to\lambda(1-g)$ of the spreading rate does not eradicate the
infection except the case $g=1$.  Indeed, by inserting Eq.~(\ref{thr})
into Eq.~(\ref{eq:gcWS}) we have that the immunization threshold is
given by
\begin{equation}
1-g_c=\frac{1}{\lambda}\frac{\avk}{\fluck}.
\label{eq:gcdef}
\end{equation}
In SF networks with $\fluck\to \infty$ only a complete immunization of the
network (i.e.  $g_c=1)$ ensures an infection-free stationary state.
The fact that uniform immunization strategies are less effective has
been noted in several cases of spatial heterogeneity
\cite{anderson92}. In SF networks we face a limiting case due to the
extremely high (virtually infinite) heterogeneity in the connectivity
properties.  Specifically, it follows from Eq.~(\ref{op}) that the SIS
model on the BA network shows for $g\simeq 1$ and any $\lambda$ the prevalence
\begin{equation}
  \rho_g\simeq 2 \exp[-1/m \lambda (1-g)].
  \label{eq:1}
\end{equation}
In other words, the infection always reaches an endemic state if the
network size is large enough (see Fig.~\ref{fig:immuno}(a)). This fact
points out the absence of an immunization threshold; SF networks are
weak in face of infections, also after massive uniform vaccination
campaigns.

\subsection{Targeted immunization}

We have seen in Section~\ref{sec:uniform-immunization} that the very
peculiar nature of SF networks hinders the efficiency of naive uniform
immunization strategies. However, we can take advantage of the 
heterogeneity of SF networks, by devising an immunization strategy 
that takes into account the inherent hierarchy in the network's nodes. 
In fact, it has been shown that SF networks posses a noticeable resilience to
\textit{random} connection
failures~\cite{barabasi00,newman00,havlin01}, which implies that the
network can resist a high level of damage (disconnected links),
without loosing its global connectivity properties; i.e. the
possibility to find a connected path between almost any two nodes in
the system.  At the same time, SF networks are strongly affected by
\textit{selective} damage; if a few of the most connected nodes are
removed, the network suffers a dramatic reduction of its ability to
carry information \cite{barabasi00,newman00,havlin01}.  Applying this
argument to the case of epidemic spreading, we can devise a
\textit{targeted} immunization scheme in which we progressively make
immune the most highly connected nodes, i.e., the ones more likely to
spread the disease. While this strategy is the simplest solution to
the optimal immunization problem in heterogeneous populations
\cite{anderson92}, its efficiency is comparable to the uniform
strategies in homogeneous networks with finite connectivity variance.
In SF networks, on the contrary, it produces an arresting increase of
the network tolerance to infections at the price of a tiny fraction of
immune individuals.

We can make an approximate calculation of the immunization threshold
in the case of a random SF network \cite{psvpro}. Let us consider the
situation in which a fraction $g$ of the individuals with the highest
connectivity have been successfully immunized.  This corresponds, in
the limit of a large network, to the introduction an upper cut-off
$k_t$---which is obviously an implicit function of the immunization
$g$---, such that all nodes with connectivity $k > k_t$ are immune.
The introduction of immune nodes implies at the same time the
elimination of all the links emanating from them, which translates, in
a mean-field approximation, into a probability $p(g)$ of deleting any
link in the network. This elimination of links yields a new
connectivity distribution, for which all moments can be computed.
Recalling Eq.~(\ref{thr}), we can then compute the critical fraction
$g_c$ of immune individuals needed to eradicate the infection. An
explicit calculation for the BA network \cite{psvpro} yields the
approximate solution for the immunization threshold in the case of
targeted immunization as
\begin{equation}
g_c\simeq \exp(-2/m\lambda).
\label{eq:gcBA}
\end{equation}
This clearly indicates that the targeted immunization program is
extremely convenient in SF networks where the critical immunization is
exponentially small in a wide range of spreading rates $\lambda$.

\begin{vchfigure}[t]
  \epsfig{file=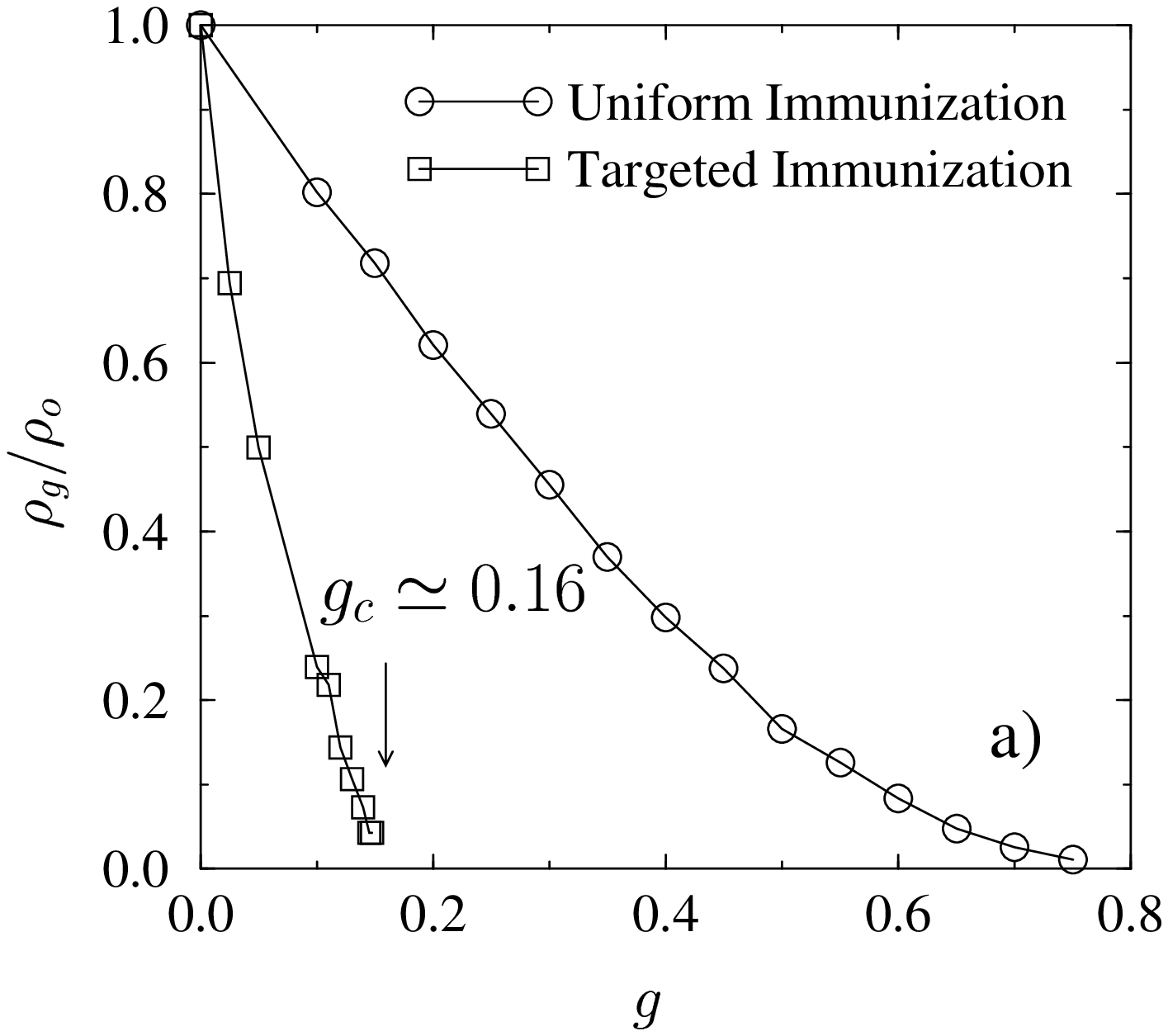, width=6cm}
    \epsfig{file=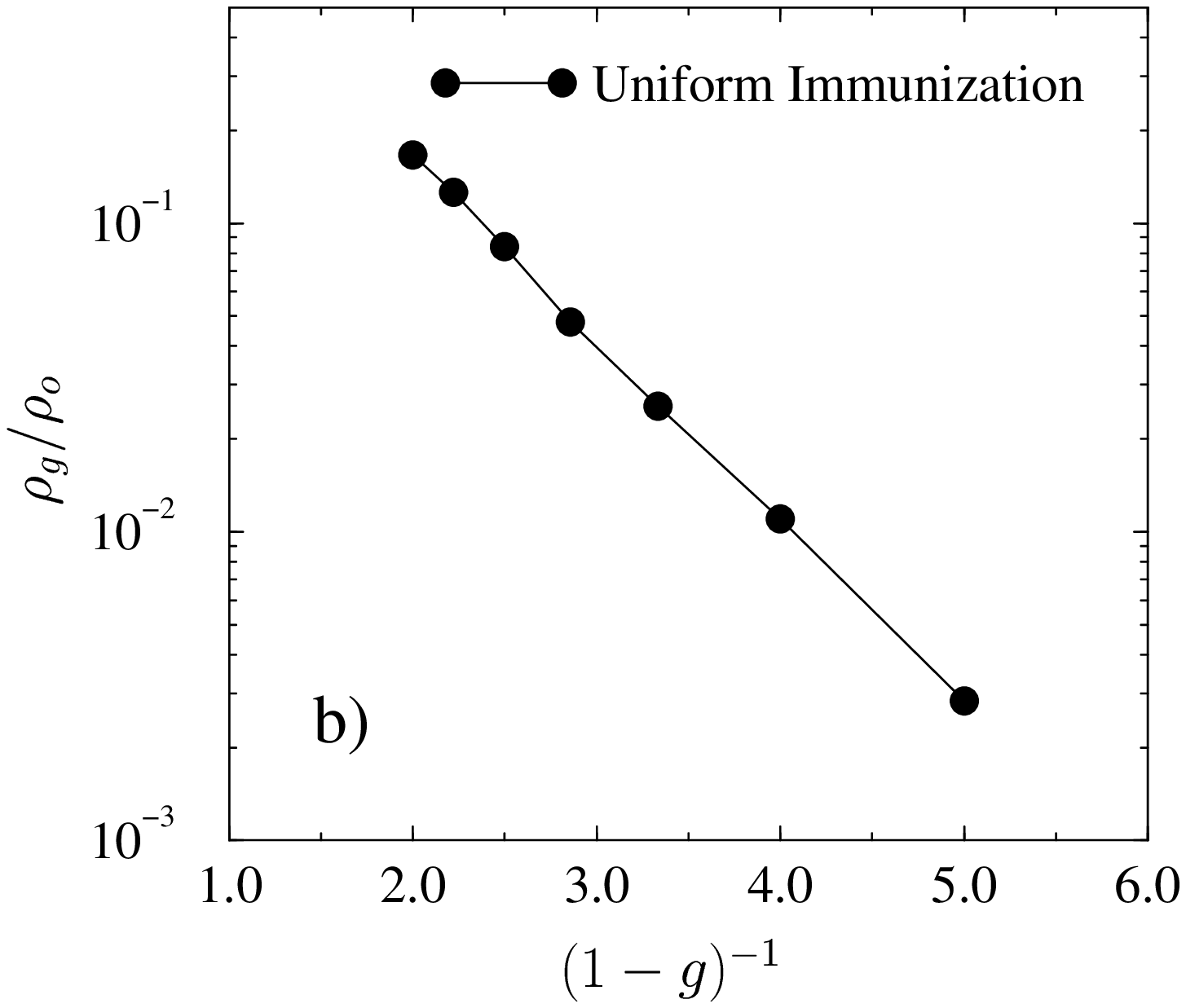, width=6cm}

\vchcaption{a) Reduced prevalence $\rho_g/ \rho_0$ from computer
    simulations of the SIS model in the BA network with uniform and
    targeted immunization, at a fixed spreading rate $\lambda=0.25$.
    A linear extrapolation from the largest values of $g$ yields an
    estimate of the threshold $g_c \simeq 0.16$ in BA networks with
    targeted immunization. %
    ~b) Check of the predicted functional dependence $\rho_g\sim
    \exp(-1/m\lambda(1-g))$ for the SIS model in the BA network with
    uniform immunization.}
  \label{fig:immuno}
\end{vchfigure}

In order to assess the efficiency of the targeted immunization scheme
we show in Fig.~\ref{fig:immuno} the results from numerical
simulations of the SIS model on BA networks, together with the results
from simulations with uniform immunization \cite{psvpro}. In
particular the plot shows the reduced prevalence $\rho_g/ \rho_0$, where
$\rho_0$ is the prevalence in the nonimmunized network, as a function of
the fraction of immunized nodes $g$, at a fixed spreading rate
$\lambda=0.25$.  Fig.~\ref{fig:immuno}(a) indicates that, for uniform
immunization, the prevalence decays slowly when increasing $g$, and
will be effectively null only for $g=1$, as predicted by
Eq.~(\ref{eq:1}). In fact, the plot in Fig.~\ref{fig:immuno}(b)
recovers the theoretical predicted behavior Eq.~(\ref{eq:1}). On the
other hand, for the targeted immunization, the prevalence shows a very
sharp drop and exhibits the onset of an immunization threshold above
which the system in infection-free. A linear regression from the
largest values of $g$ yields an approximate estimation $g_c\simeq0.16$,
that definitely proves that SF networks are very sensitive to the
targeted immunization of a very small fraction of the most connected
nodes.

This result can be readily extended to SF networks with arbitrary $\gamma$
values, and it is also possible to devise alternative immunization
schemes which take advantage of the SF connectivity patterns in order
to achieve a high level of tolerance to infections~\cite{psvpro}.
Other strategies have been put forward in Ref.~\cite{aidsbar} by
proposing to cure with proportionally higher rates the most connected
nodes.  Also in this case it is possible to reintroduce a threshold in
the network with this hub-biased policy for the administered cures.

The present results indicate that the SF networks' susceptibility to
epidemic spreading is reflected also in an intrinsic difficulty in
protecting them with local---uniform---immunization. On a global
level, uniform immunization policies are not satisfactory and only
targeted immunization strategies successfully lower the vulnerability
of SF networks.  This evidence radically changes the usual perspective
of the regular epidemiological framework. Spreading of infectious or
polluting agents on SF networks, such as food or social webs, might be
contrasted only by a careful choice of the immunization procedure. In
particular, these procedures should rely on the identification of the
most connected individuals. The protection of just a tiny fraction of
these individuals raises dramatically the tolerance to infections of
the whole population.  The computer virus case is once again providing
support to this picture.  Despite deployment of antivirus software is
timely and capillary, viruses' lifetimes are extremely long; in other
words, very high levels of immunization are not able to eradicate the
epidemic. In the standard epidemic framework this would be possible
only in the case of very high spreading rate for the virus that is in
contradiction with the always small prevalence of epidemic outbreaks.
These empirical findings are, however, in good agreement with the
picture obtained for the immunization of SF networks.  In fact, the
antivirus deployment is not eradicating the epidemics on the global
scale since it is alike to a random immunization process where file
scanning and antivirus updating are statistically left to the good
will of users and system managers. Needless to say, from the point of
view of the single user, antiviruses are extremely important, being
the only way to ensure local protection for the computer.

\section{Conclusions}

The topology of the network has a great influence in the overall
behavior of epidemic spreading. The connectivity fluctuations of the
network play a major role by strongly enhancing the infection's
incidence.  This issue assumes a particular relevance in the case of
SF networks that exhibit connectivity fluctuations diverging with the
increasing size $N$ of the web.  Here we have reviewed the new
epidemiological framework obtained in population networks
characterized by a scale-free connectivity pattern. SF networks are
very weak in face of infections, presenting an effective
epidemic threshold that is vanishing in the limit $N\to \infty$. In an
infinite population this corresponds to the absence of any epidemic
threshold below which major epidemic outbreaks are impossible. SF
networks' susceptibility to epidemic spreading is reflected also in an
intrinsic difficulty in protecting them with
local---uniform---immunization policies. Only targeted immunization
procedures achieve the desired lowering of epidemic outbreaks and
prevalence.

The present picture qualitatively fits the observations from real data
of computer virus spreading, and could solve the long standing problem
of the generalized low prevalence and long lifetime of computer
viruses without assuming any global tuning of the spreading rates.
Moreover, recent findings on the web of human sexual contacts~\cite{amaral01}
prompt that the presented results could have potentially interesting
implications also in the case of human sexual disease control.

In order to illustrate the new features of epidemic spreading in SF
networks, we used the SIS model. It is important to stress, however,
that the analysis on SF networks of different models, such as the SIR
model, confirm the presented epidemiological
picture~\cite{lloydsir,moreno,newman02}.  Yet, many other ingredients
concerning the infection mechanisms should be considered in a more
realistic representation of real epidemics
\cite{anderson92,epidemics}. In addition, simple rules defining the
temporal patterns of the networks, such as the frequency of forming
new connections, the actual time that a connection exists,
or different types of connections, should be included in the modeling.
These dynamical features are highly valuable experimental inputs which
are necessary ingredients in the use of complex networks theory in
epidemic modeling.

\section*{Acknowledgements}

This work has been partially supported by the European Network
Contract No. ERBFMRXCT980183 and by the European Commission - Fet Open
project COSIN IST-2001-33555.  R.P.-S. acknowledges financial support
from the Ministerio de Ciencia y Tecnolog\'{\i}a (Spain).

\renewcommand\bibname{References}

\end{document}